\documentclass[10pt,english,british,tightenlines,eqsecnum,
floats,aps,amsmath,amssymb,nofootinbib,superscriptaddress,prd,showpacs,showkeys]
              {revtex4}
\usepackage[latin9]{inputenc}
\usepackage{color}
\usepackage{babel}
\usepackage{amsmath}
\usepackage{amssymb}
\usepackage{graphicx}
\usepackage{esint}
\usepackage[unicode=true,pdfusetitle,
 bookmarks=true,bookmarksnumbered=false,bookmarksopen=false,
 breaklinks=false,pdfborder={0 0 1},backref=false,colorlinks=false]
 {hyperref}
\makeatletter
\@ifundefined{textcolor}{}
{%
 \definecolor{BLACK}{gray}{0}
 \definecolor{WHITE}{gray}{1}
 \definecolor{RED}{rgb}{1,0,0}
 \definecolor{GREEN}{rgb}{0,1,0}
 \definecolor{BLUE}{rgb}{0,0,1}
 \definecolor{CYAN}{cmyk}{1,0,0,0}
 \definecolor{MAGENTA}{cmyk}{0,1,0,0}
 \definecolor{YELLOW}{cmyk}{0,0,1,0}
}

\@ifundefined{date}{}{\date{01-March-2022}}
\usepackage{babel}

\usepackage{euscript}\usepackage{epsfig}

\usepackage{amsfonts}

\usepackage{fancyhdr}

\makeatother

\begin{document}

\title{Noncommutativity, S\'{a}ez-Ballester theory and kinetic inflation}
\author{S. M. M. Rasouli}

\email{mrasouli@ubi.pt}

\affiliation{Departamento de F\'{i}sica,
Centro de Matem\'{a}tica e Aplica\c{c}\~{o}es (CMA-UBI),
Universidade da Beira Interior,
Rua Marqu\^{e}s d'Avila
e Bolama, 6200-001 Covilh\~{a}, Portugal;
}

 \affiliation{Department of Physics, Qazvin Branch, Islamic Azad University, Qazvin, Iran}

\begin{abstract}

This paper presents a noncommutative (NC)
version of an extended S\'{a}ez-Ballester (SB) theory. Concretely, considering the spatially flat
Friedmann-Lema\^{\i}tre-Robertson-Walker~(FLRW) metric,
we propose an appropriate
dynamical deformation between the conjugate
momenta and applying the Hamiltonian formalism, obtain deformed equations of motion.
In our model, the NC parameter appears linearly in
the deformed Poisson bracket and the equations of
the NC SB cosmology. When it goes
to zero, we get the corresponding commutative counterparts.
Even by restricting our attention to a particular case,
where there is neither an ordinary matter nor a scalar
potential, we show that the effects of the noncommutativity provide interesting results:
applying numerical endeavors for very small values of the NC parameter, we show that (i) at the early
times of the universe, there is an inflationary
phase with a graceful exit, for which the relevant nominal condition is satisfied;
(ii) for the late times, there is a zero acceleration epoch.
By establishing an appropriate dynamical
framework, we show that the results (i) and (ii) can be obtained
for many sets of the initial conditions and the parameters of the model.
Finally, we indicate that, at the level of the field equations, one
may find a close resemblance between our NC model and the Starobinsky inflationary model.
 \end{abstract}

% Keywords

% The fields PACS, MSC, and JEL may be left empty or commented out if not applicable
%\PACS{J0101}
%\MSC{}
%\JEL{}

%%%%%%%%%%%%%%%%%%%%%%%%%%%%%%%%%%%%%%%%%%
% Only for the journal Diversity
%\LSID{\url{http://}}

%%%%%%%%%%%%%%%%%%%%%%%%%%%%%%%%%%%%%%%%%%
% Only for the journal Applied Sciences:
%\featuredapplication{Authors are encouraged to provide a concise description of the specific application or a potential application of the work. This section is not mandatory.}
%%%%%%%%%%%%%%%%%%%%%%%%%%%%%%%%%%%%%%%%%%

%%%%%%%%%%%%%%%%%%%%%%%%%%%%%%%%%%%%%%%%%%
% Only for the journal Data:
%\dataset{DOI number or link to the deposited data set in cases where the data set is published or set to be published separately. If the data set is submitted and will be published as a supplement to this paper in the journal Data, this field will be filled by the editors of the journal. In this case, please make sure to submit the data set as a supplement when entering your manuscript into our manuscript editorial system.}

%\datasetlicense{license under which the data set is made available (CC0, CC-BY, CC-BY-SA, CC-BY-NC, etc.)}

%%%%%%%%%%%%%%%%%%%%%%%%%%%%%%%%%%%%%%%%%%
% Only for the journal Toxins
%\keycontribution{The breakthroughs or highlights of the manuscript. Authors can write one or two sentences to describe the most important part of the paper.}

%%%%%%%%%%%%%%%%%%%%%%%%%%%%%%%%%%%%%%%%%%
% Only for the journal Encyclopedia
%\encyclopediadef{Instead of the abstract}
%\entrylink{The Link to this entry published on the encyclopedia platform.}
%%%%%%%%%%%%%%%%%%%%%%%%%%%%%%%%%%%%%%%%%%

\maketitle
\section{Introduction}
\label{int}
\indent

To overcome the problems of standard
cosmology, various alternative theories
to general relativity have been established.
Among them, the scalar-tensor theories have played a significant role, see, for
instance, \cite{Faraoni.book,DDB07, Q19,K19,Book-S21} and reference therein.
In the S\'{a}ez-Ballester (SB) scalar-tensor theory \cite{SB85-original}, in which the scalar
field is minimally coupled to gravity, a particular non-canonical kinetic term was added to
the Einstein-Hilbert action. The original Lagrangian associated with the SB theory
includes the ordinary matter sector, but neither cosmological constant nor
a scalar potential have contributed to it.
Moreover, we should mention that the SB theory possesses
 dimensionless parameters $n$ and ${\cal W}$, in which the latter specifies
the strength of the coupling between the gravity and the
SB scalar field. To the best of our knowledge, it has not
been investigated for which values of ${\cal W}$, the observational limits can be satisfied.
The SB theory and its extended versions, in both the
classical and quantum levels, have been widely applied to investigate
various cosmological problems in either four or arbitrary
dimensions \cite{P87,SA91,SS03,MSM07, SSL10,NSR12,JAMM12,Y13,RPR15,RKN12,RM18,RPSM20}.
Another category of alternative theories has arisen due to the incapability of the GR
in predicting the effects of some phenomena at the Planck regime \cite{DV08,B15}.
Among such theories, one can refer to some approaches to noncommutative (NC)
gravity (see, \cite{CORS03,GS06,HR06,JCOR08,Add21},
and references therein), which has roots in noncommutative
geometry and noncommutative quantum field theories.
As these frameworks are highly nonlinear, therefore, for investigating
effects of noncommutativity on different aspects of the
universe, noncommutative cosmology has been proposed, see, for instance, \cite{COR02,BN04, PM05}.
It has been believed that for constructing
 noncommutative models, both at the quantum as well
as classical regime, cosmology can be considered as an interesting arena \cite{PSR13}.
For instance, at the classical regime, by modifying the Poisson
brackets of the classical theories, one can obtain noncommutative equations of motion.
In these frameworks, by including a NC parameter
which is usually interpreted as the Planck (length) constant, the effects of the noncommutativity
may assist to resolve a few open problems of
cosmology \cite{RFK11,GSS11,RM14,LSY18, RMM19,OM21,MM21}.

The main objective of the present work is
to establish a noncommutative cosmological model in the
context of the SB scalar-tensor theory containing
an arbitrary potential.
Subsequently, we will study the effects of noncommutativity in a
 particular case where the ordinary matter, as well as
scalar potential, are absent. For such a simple model, we will see that in addition to
the NC parameter, the presence of $n$ and the SB
coupling parameter ${\cal W}$ are also significant in describing the universe at an early time.
Moreover, it is worthy to mention that, if the noncommutativity is
present at a small scale, by the UV/IR mixing (which is a feature of the noncommutativity),
it can also be observed at late times of the universe.

The paper is outlined as follows.
In the next section, considering a spatially flat Friedmann-Lema\^{\i}tre-Robertson-Walker~(FLRW) metric as the background
 geometry and applying the Hamiltonian approach, we will obtain
cosmological equations of motion for an extended SB cosmology in the non-deformed case. Then, we propose
a general dynamical deformation (noncommutativity) to establish an interesting cosmological scenario.
%In this work, for convenience and simplicity, we focus on a particular case where $n=0$.
%Such a simple case, we can also observe the effects of the coupling parameter ${\cal W}$ with more care.
In section \ref{free potential}, we first obtain analytic exact cosmological solutions
 for the commutative model. Subsequently, we focus on the NC model and
 show that, for very small values of the NC parameter,
 there is an inflationary phase, with graceful exit, at the early time.
 Moreover, we show that, for our herein NC model, the nominal
 condition associated with the inflation is satisfied.
 Furthermore, our numerical endeavors show that the
 noncommutative effects can also be seen at late times.
 Concretely, for the latter, we observe that the scale factor increase with zero acceleration.
 In section \ref{dyn}, by proposing an appropriate dynamical
 setting, we show that the above-mentioned results are confirmed.
Finally, we present our conclusions in section~\ref{concl}.

\section{A noncommutative cosmological scenario in S\'{a}ez-Ballester theory}
\label{Standard}

We start with the spatially flat FLRW universe
\begin{equation}\label{frw}
ds^{2}=-\mathcal{N}^2(t)dt^2+a^2(t)\left(dx^2+dy^2+dz^2\right),
\end{equation}
where $t$ is the cosmic time, $x, y, z$ are the Cartesian
coordinates, $a(t)$ is the scale
factor and $\mathcal{N}(t)$ is a lapse function.
Let us consider an extended
version of the SB Lagrangian density:
\begin{equation}
{\mathcal{L}}=\sqrt{-g}\,\left[R-{\cal W}\phi^n\, g^{\alpha\beta}\,({\cal
D}_\alpha\phi)({\cal D}_\beta\phi)-V(\phi)+\chi\,L_{_{\rm matt}}\right],
\label{eq1}
\end{equation}
where $\chi\equiv 8\pi G$; $n$ and ${\cal W}$ are dimensionless
{\it independent} parameters, $g$ denotes the determinant of the metric
$g_{\mu\nu}$, $R$ is the Ricci scalar, the Greek indices run from zero to three and we have assumed the
 units where $c=1=\hbar$. The scalar field $\phi$ is minimally
coupled to gravity, $V(\phi)$ is a scalar potential, $L_{_{\rm matt}}=-2\varrho(a)$ is the Lagrangian
density associated with the ordinary matter and ${\cal
D}_\alpha$ denotes the covariant derivative.

Substituting the Ricci scalar associated with the metric \eqref{frw} into \eqref{eq1}, we obtain
\begin{equation}\label{Lag}
{\mathcal{L}}=-6\mathcal{N}^{-1}a\dot{a}^2+{\cal W}\mathcal{N}^{-1}a^3{\phi}^n\dot{\phi}^2
-\mathcal{N}a^3\left[2\chi\varrho(a)+V(\phi)\right],
\end{equation}
where a dot denotes a derivative with respect to the time and we have neglected a total time derivative term.
It is straightforward to show that the Hamiltonian of the model is given by
\begin{equation}\label{H0}
\mathcal{H}=-\frac{\mathcal{N}}{24}a^{-1}P_{a}^{2}
+\frac{\mathcal{N}}{4{\cal W}}a^{-3}{\phi}^{-n}P_{\phi}^{2}+\mathcal{N}a^3\left[2\chi\varrho(a)+V(\phi)\right],
\end{equation}
where $P_a$ and $P_{\phi}$ stand for the momenta conjugates of the scale factor and the scalar field,
respectively.

Considering the comoving gauge, i.e, setting $\mathcal{N}=1$, employing the
Hamiltonian \eqref{H0}, and admitting the Poisson algebra
$\{a,\phi\}=0$, $\{P_a,P_\phi\}=0$,
$\{a,P_a\}=1$ and $\{\phi,P_\phi\}=1$ for the phase space coordinates
$\{a,\phi;P_a,P_\phi\}$,
we easily obtain:
\begin{eqnarray}\label{diff.eq1}
\dot{a}\!\!&=&\!\!-\frac{1}{12}a^{-1}P_a\,,\\
\label{diff.eq2}
\dot{P_a}\!\!&=&\!\!-\frac{1}{24}a^{-2}P_a^2+\frac{3}{4{\cal W}}a^{-4}{\phi}^{-n}P_\phi^2-3a^2\left(2\chi\varrho+V\right)-2\chi a^3\frac{d\varrho(a)}{da},
\\
\label{diff.eq3}
\dot{\phi}\!\!&=&\!\!\frac{1}{2{\cal W}}a^{-3}{\phi}^{-n}P_\phi\,,\\
 \label{diff.eq4}
\dot{P_\phi}\!\!&=&\!\!\frac{n}{4{\cal W}}a^{-3}{\phi}^{-(n+1)}P_\phi^2- a^3\frac{dV(\phi)}{d\phi}.
  \end{eqnarray}
Using equations \eqref{diff.eq1}-\eqref{diff.eq4}, it is straightforward to obtain the
equations of the non-deformed SB cosmological model:
%For such a NC setting, it is easy to show that the equations of motion are give by
\begin{eqnarray}\label{C1}
3 H^{2}=\frac{1}{2}\left[{\cal W}{\phi}^{n}\dot{\phi}^{2}+V(\phi)\right]+\chi\varrho(a),
\end{eqnarray}
\begin{eqnarray}\label{C2}
2\frac{\ddot{a}}{a}+H^{2}=-\frac{1}{2}\left[{\cal W}{\phi}^{n}\dot{\phi}^{2}-V(\phi)\right]+\chi\left[\varrho(a)+\frac{a}{3}\frac{d\varrho(a)}{da}\right],
\end{eqnarray}
\begin{equation}\label{C3}
\ddot{\phi}+3H\dot{\phi}+\frac{1}{2{\cal W}}{\phi}^{-n}\frac{dV(\phi)}{d\phi}+\frac{n}{2}\phi^{-1}\dot{\phi}^2=0,
\end{equation}
where $H\equiv\dot{a}/a$ is the Hubble parameter.
\\
In order to establish an appropriate noncommutative scenario, we would
propose a (dynamical) deformation solely between the
conjugate momenta as\footnote{Some arguments for such a deformation have been presented in \cite{RFK11}.}
\begin{equation}\label{deformed}
\{P_{a},P_{\phi}\}=\theta\phi^{2n+3},
\end{equation}
(where $\theta={\rm constant}$ is the NC parameter)
and leave the other Poisson brackets unchanged.

Under the NC deformation \eqref{deformed}, the equations
\eqref{diff.eq1} and \eqref{diff.eq3} remain unchanged. However, the equations associated with the
momenta, i.e., equations \eqref{diff.eq2} and \eqref{diff.eq4}, are deformed:\footnote{In what follows, let us briefly
 present another approach to obtain the NC field equations, see, for instance, \cite{DS04,RZMM14}.
 In order to obtain the Hamiltonian corresponding to the NC model, we proceed as follows.
 (i) All the variables of \eqref{H0} should be replaced by new ones, for instance, primed variables.
 (ii) Introducing the only transformation $P'_\phi=P_\phi-\theta a\phi^{2n+3}$, and assuming
 that the other primed variables are equal to the corresponding unprimed
 ones, we can easily recover not only the deformed
 Poisson bracket \eqref{deformed} but also the NC Hamiltonian.
 (iii) Finally, using the latter together with usual
 (standard) Poisson brackets, we can easily obtain the NC
 counterparts of \eqref{diff.eq1}-\eqref{diff.eq4}.}
\begin{eqnarray}\label{diff.eq2-dyn}
\dot{P_a}\!\!&=&\!\!-\frac{1}{24}a^{-2}P_a^2+\frac{3}{4{\cal W}}a^{-4}{\phi}^{-n}P_\phi^2-3a^2\left(2\chi\varrho+V\right)-2\chi a^3\frac{d\varrho(a)}{da}\cr\cr
&+&\frac{\theta}{2{\cal W}}a^{-3}{\phi}^{n+3}P_\phi,\\\nonumber
\\
 \label{diff.eq4-dyn}
\dot{P_\phi}\!\!&=&\!\!\frac{n}{4{\cal W}}a^{-3}{\phi}^{-(n+1)}P_\phi^2- a^3\frac{dV(\phi)}{d\phi}+\frac{\theta}{12}a^{-1}{\phi}^{2n+3}P_a,
  \end{eqnarray}
where we have used
\begin{eqnarray}\label{calc}
\{P_a,h(P_a,P_\phi)\}=\theta\phi^{2n+3}\frac{\partial h}{\partial P_\phi},\\
\{P_\phi,h(P_a,P_\phi)\}=-\theta\phi^{2n+3}\frac{\partial h}{\partial P_a},
\end{eqnarray}
in which $h$ is an arbitrary function of the conjugate momenta.

 It is easy to show that the equations of motion associated with
 our herein NC framework are given by
\begin{eqnarray}\label{asli1}
3 H^{2}=\frac{1}{2}\left[{\cal W}{\phi}^{n}\dot{\phi}^{2}+V(\phi)\right]+\chi\varrho(a),
\end{eqnarray}
\begin{eqnarray}\label{asli2}
2\frac{\ddot{a}}{a}+H^{2}=-\frac{1}{2}\left[{\cal W}{\phi}^{n}\dot{\phi}^{2}-V(\phi)\right]+\chi\left[\varrho(a)+\frac{a}{3}\frac{d\varrho(a)}{da}\right]-\frac{\theta}{6}a^{-2} \phi^{2n+3}\dot{\phi},
\end{eqnarray}
\begin{equation}\label{asli3}
\ddot{\phi}+3H\dot{\phi}+\frac{n}{2}{\phi}^{-1}\dot{\phi}^{2}+\frac{1}{2{\cal W}}{\phi}^{-n}\frac{dV(\phi)}{d\phi}+\frac{\theta}{2{\cal W}} a^{-2}H\phi^{n+3}=0.
\end{equation}

%where we have focused only on a simple case where $n=0$.
We should note that in a particular case, where the NC parameter $\theta$ vanishes, equations
\eqref{asli1}-\eqref{asli3} reduce to their non-deformed counterparts.
In the subsequent sections, we restrict our attention to a specific case of a formerly constructed NC framework.

\section{Kinetic inflation and the horizon problem}
\label{free potential}
In this section, let us investigate a very simple set up of our herein NC and commutative
models, in which the ordinary
matter, as well as the scalar potential, are absent, i.e., $\varrho=0$ and $V(\phi)=0$.
Therefore, the energy density and pressure reduce to
\begin{eqnarray}\label{rho-tot}
\rho_{_{\rm tot}}&=&\frac{{\cal W}}{2}{\phi}^{n}\dot{\phi}^{2},\\\nonumber\\
\label{p-tot}
p_{_{\rm tot}}&=&\frac{{\cal W}}{2}{\phi}^{n}\dot{\phi}^{2}+p_{_{\rm nc}}, \hspace{10mm}
p_{_{\rm nc}}\equiv\frac{\theta}{6}a^{-2} \phi^{2n+3}\dot{\phi}.
\end{eqnarray}
Using equations \eqref{asli3}-\eqref{p-tot}, it is straightforward to show that the
conservation equation for the matter (associated with the NC framework) is identically satisfied:
\begin{eqnarray}\label{cons-eq}
\dot{\rho}_{_{\rm tot}}+3H\left(\rho_{_{\rm tot}}+p_{_{\rm tot}}\right)=0.
\end{eqnarray}

It is important to note that $p_{_{\rm nc}}$, in equation \eqref{p-tot},
depends {\it explicitly} on the NC parameter $\theta$. Although, the NC
parameter does not, explicitly, appear in the relations
associated with the $\rho_{_{\rm tot}}$ (or equivalently, $\rho_{_{\rm tot}}-p_{_{\rm nc}}$), but we
will show that they also depend on the NC parameter, {\it implicitly}.
The above statements point out that there is no way
to specify the commutative sector of any
quantity, unless
substituting $\theta=0$ in all the equations of motion.

Substituting $\varrho=0=V$ into \eqref{asli1}, we
obtain
\begin{equation}\label{exp-rel-2}
H=\alpha {\phi}^{\frac{n}{2}}\dot{\phi}(t), \hspace{10mm}\alpha\equiv \pm\sqrt{\frac{{\cal W}}{6}},
\end{equation}
where we have assumed ${\cal W}>0$.
Moreover, assuming $n\neq-2$, from using equation \eqref{exp-rel-2}, we
obtain the scale factor as a function of the SB scalar field as\footnote{In this work, let us skip the cosmological model corresponding to $n=-2$, see section \ref{concl}.}
\begin{equation}\label{Ex1-Vphi}
a=a_i\exp \left[f(\phi)\right],\hspace{8mm}
f(\phi)\equiv\frac{2\alpha}{n+2}\left(\phi^{\frac{n+2}{2}}-\phi_i^{\frac{n+2}{2}}\right),
 \end{equation}
where $a_i$ and $\phi_i$ are integration constants.
By substituting \eqref{exp-rel-2} and \eqref{Ex1-Vphi} into \eqref{asli3}, the wave equation
will be a differential equation for the scalar field only:
 \begin{eqnarray}\label{c-zero-V-1}
\ddot{\phi}\!\!&+&\!\!\left(3\alpha{\phi}^{\frac{n}{2}}+\frac{n}{2}{\phi}^{-1}\right)\dot{\phi}^2
+ \left(\frac{\theta}{12\alpha a_i^2}\right)\phi^{\frac{3}{2}(n+2)}\exp \left[-2f(\phi)\right]\dot{\phi}=0.
\end{eqnarray}
In what follows, we are going to solve equation \eqref{c-zero-V-1} either analytically or
numerically, by which we will present analysis for the commutative and NC cosmological models.
\\

\subsection{Commutative case}
\label{CC}
Let us first investigate the commutative case, which will be
required later to compare with the corresponding NC case.
Substituting $\theta=0$ into equation \eqref{c-zero-V-1}, we can easily obtain an exact solution:
\begin{equation}\label{commut-phi}
\phi (t)=\left\{\frac{(n+2) \ln \left[3 \alpha  c_1 (t+c_2)\right]}{6 \alpha }\right\}^{\frac{2}{n+2}},
\end{equation}
where $c_1$ and $c_2$ are integration constants.
Moreover, substituting \eqref{commut-phi} into the corresponding
relation of scale factor in \eqref{Ex1-Vphi}, we obtain
\begin{equation}\label{commut-a}
a(t)=a_i \exp\left[-\frac{2\alpha}{n+2}\phi_i^{\frac{2}{n+2}}\right]\left[3 \alpha  c_1 (t+c_2)\right]^{\frac{1}{3}},
 \end{equation}
which implies that, in the commutative case, the scale factor of the universe decelerates forever.
 In a particular case where $n=0$ and ${\cal W}=1$, the solutions \eqref{commut-phi} and \eqref{commut-a}
 reduce to the corresponding ones obtained in \cite{SS17}.
%Disregarding the values, which can be taken by the SB coupling
%parameter, one can read the discussions presented for this case in \cite{SS17}.
Let us abstain from analyzing these results here.
However, the behavior of the physical quantities for
this case will be described and compared with the corresponding NC model, see, for
instance, figures \ref{phi-zero}-\ref{PhaseP}.

\subsection{Noncommutative case}
\label{NC case}

For the NC case, i.e., $\theta\neq0$, it is not feasible to obtain
exact solutions for complicated differential
equation \eqref{c-zero-V-1}, analytically.
In this respect, let us investigate this case by applying numerical methods.
Concretely, assuming $n\neq-2$, we apply the numerical solution of
the equation \eqref{c-zero-V-1} to plot the physical quantities.
In what follows, we will present briefly the consequences
of our numerical endeavors, which have been obtained for every proper
set of the initial conditions (ICs), the values of the parameters of the model, and the integration constants.
We should note that for every set, we have taken
very small negative values of the NC parameter.
Our results are (see, for instance, figures \ref{phi-zero}-\ref{Dh}):
\begin{enumerate}
  \item
  In the early times, both the scalar field and the
  scale factor experience accelerated expansion.

  \item
   Thereafter, there is another different phase in which both of them decelerate, see figures \ref{phi-zero}.

  \item
   At late times, both $\ddot{\phi}$ and $\ddot{a}$ asymptotically
   tend to zero.

   Up to now, we can conclude that stages 1 and 2 indicate that our NC model may be
 considered as a successful cosmological inflationary
 model (see also the discussions will be presented in the following).
 Concretely, an inflationary phase took place at the earlier times, and
 afterward, there is a radiation-dominated epoch. Moreover,
 the effects of the dynamical noncommutativity \eqref{deformed} provide an
 appropriate transition from the accelerating phase to the
 decelerating one, which is known as the graceful exit.
 However, stage (iii) may be interpreted as a quantum
gravity footprint as the coarse-grained explanation.

 \item
  Let us now analyze the time behavior of the energy density and
pressure, see, for instance, the upper right panel of figure \ref{rho-p}.
It is seen that $\rho_{_{\rm tot}}$ always takes positive
values such that it increases during the inflationary epoch to reach its maximum value.
Soon after exiting from the accelerated phase, it decreases forever. Whilst, both the
$p_{_{\rm tot}}$ and $p_{_{\rm nc}}$ always take negative values.
In contrast to the $\rho_{_{\rm tot}}$, they decrease during the inflationary phase
whilst increasing during the radiation-dominated era.
They reach their minimum value at the moment of the transition phase.

 \item
  In contrast to exact solutions, we should not always expect a numerical
  solution to satisfy the conservation equation identically. In this regard, it is worth plotting the quantity on
the left-hand side of equation \eqref{cons-eq} to find out how much
 disperses from zero (for this we use the numerical solution of equation \eqref{c-zero-V-1}).
  Therefore, for every set of the ICs and the values of
the parameters, which have been used to depict the behavior
of the quantities, we have checked its corresponding degree of accuracy.
Specifically, for every numerical set, we have
plotted the corresponding numerical error to be sure
that they whether or not satisfy the conservation
equation \eqref{cons-eq}, see, for instance, the lower panels of figure \ref{rho-p}.

 \item
  We have also investigated the time behavior of $\phi(t)$, $a(t)$, their
first and second derivatives (with respect to the cosmic time)
for different values of the parameters ${\cal W}$ and $n$, see, for
 instance, figures \ref{phi-diff-W} and \ref{phi-diff-n}, which
 show the behavior of $\phi(t)$ and $a(t)$ against the
cosmic time. Our consequences
indicate that, for a specific set of values,
by changing the values of ${\cal W}$ (or $n$) and leaving the
others unchanged, there
are no perceptible changes in the general
behavior of the quantities, which was reported by in stages 1 to 5.
Notwithstanding, we found that for any $t$, assuming ${\cal W}>0$, the smaller the
value of ${\cal W}$, the larger the values of $a$ and $\phi$.
Moreover, our endeavors have shown that the smaller the value of ${\cal W}$, the
shorter the amount of the interval time of the inflationary epoch.
According to figure \ref{phi-diff-n}, an
interpretation can also be presented for the case if only $n$ varies.

\begin{figure}
\includegraphics[width=3in]{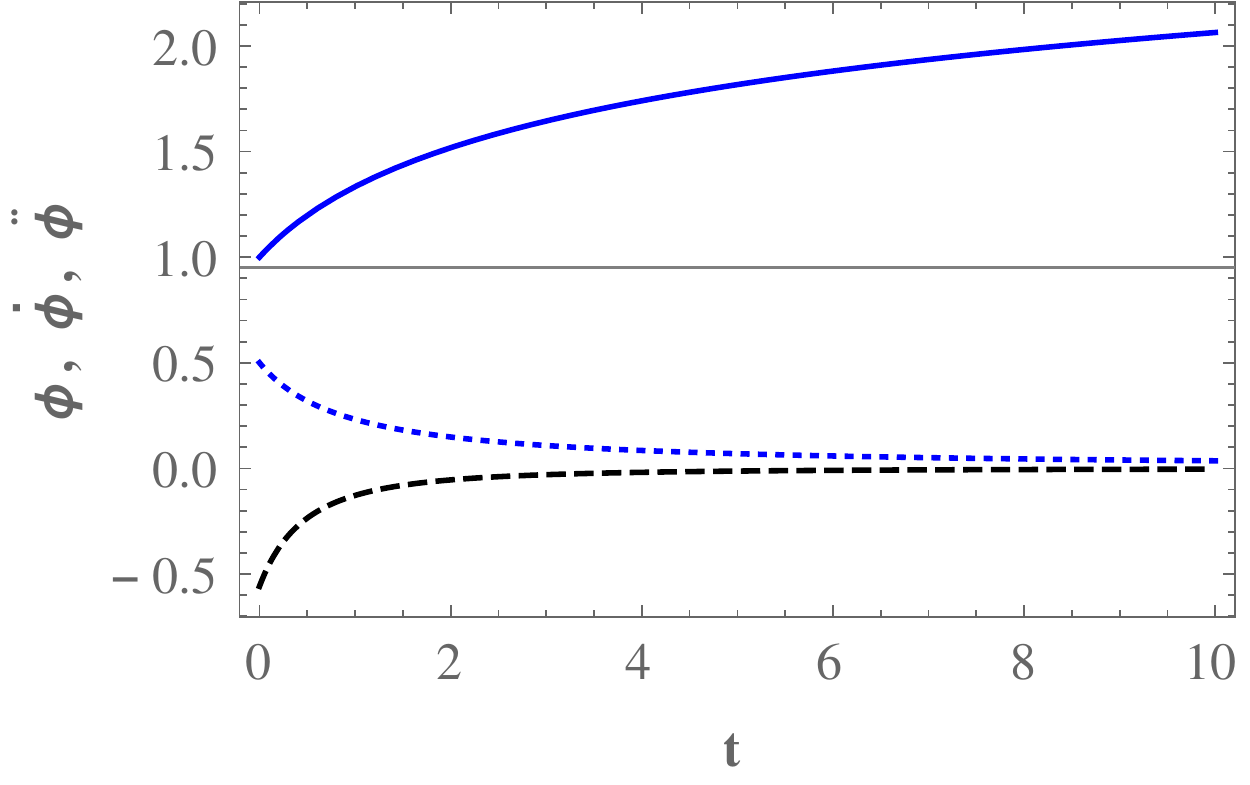}
\hspace{2mm}
\includegraphics[width=3in]{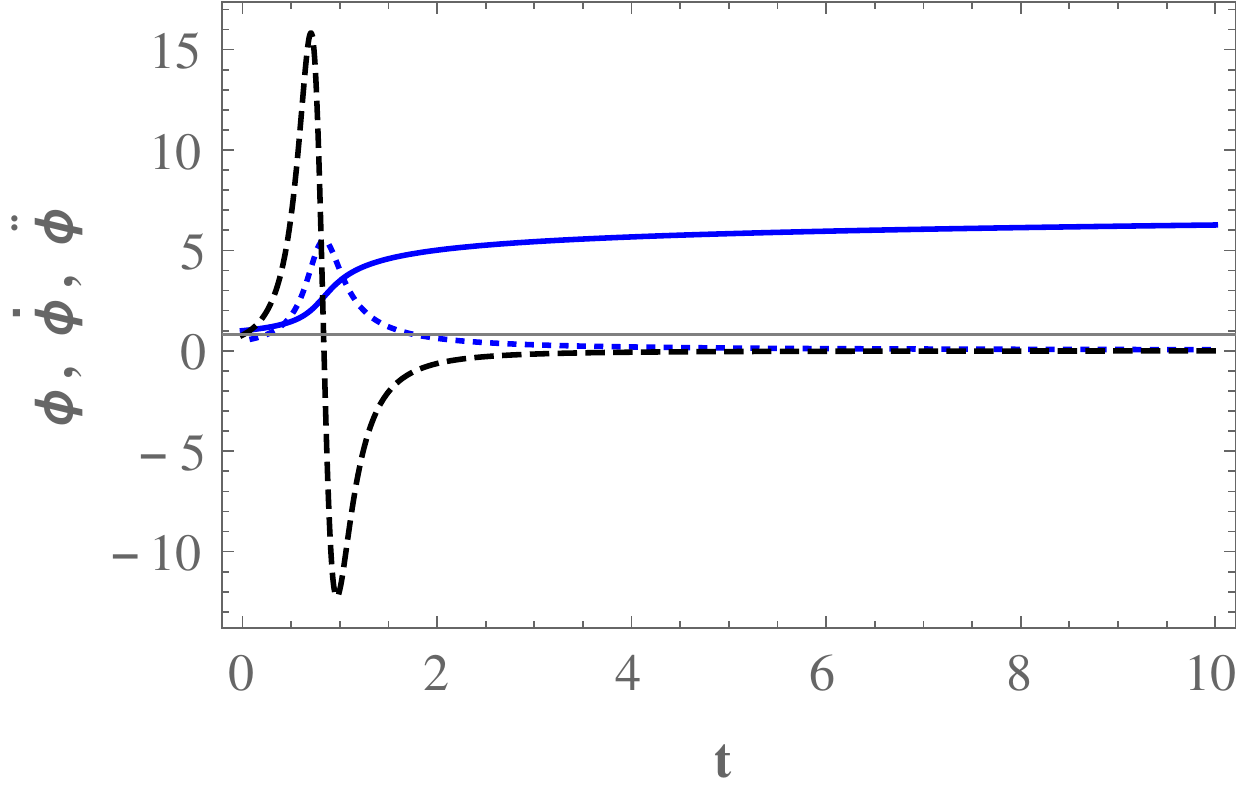}
\hspace{2mm}
\includegraphics[width=3in]{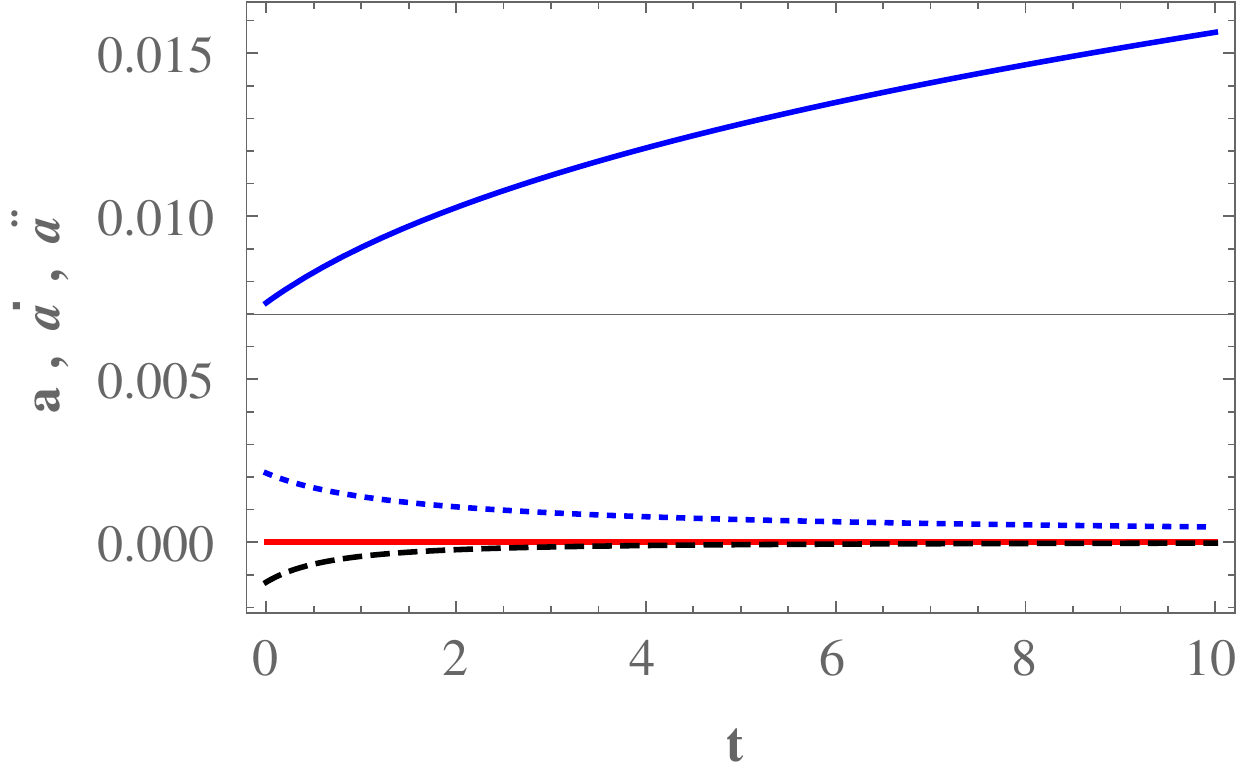}
\hspace{2mm}
\includegraphics[width=3in]{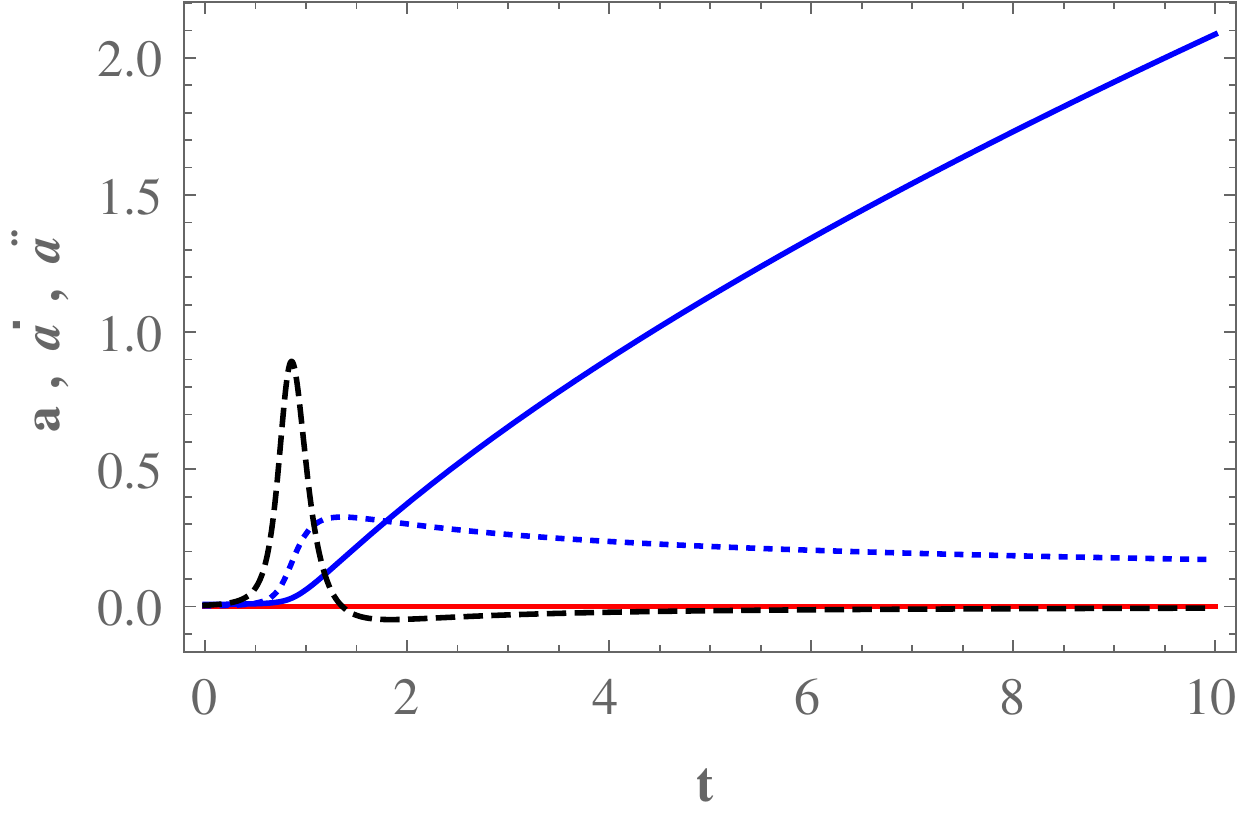}
\caption{The behavior of the scalar field $\phi(t)$ and the scale factor $a(t)$ (solid curves),
their first and second derivatives (dotted and dashed curves, respectively)
 against cosmic time for the commutative model, for which $\theta=0$, (the left panels) and
 NC model with $\theta=-0.001$ (the right panels).
Moreover, for both the commutative and NC models, we have
set $n=1$, ${\cal W}=2$, $8\pi G=1$, $a_i=0.005$, $\phi_i=0.01$, $\phi(0)=1$, $\dot{\phi}(0)=0.5$.
 To recognize the quantities with positive and negative values, we have plotted $a(t)=0$ with a red line.}
\label{phi-zero}
\end{figure}

\begin{figure}
\centering\includegraphics[width=3in]{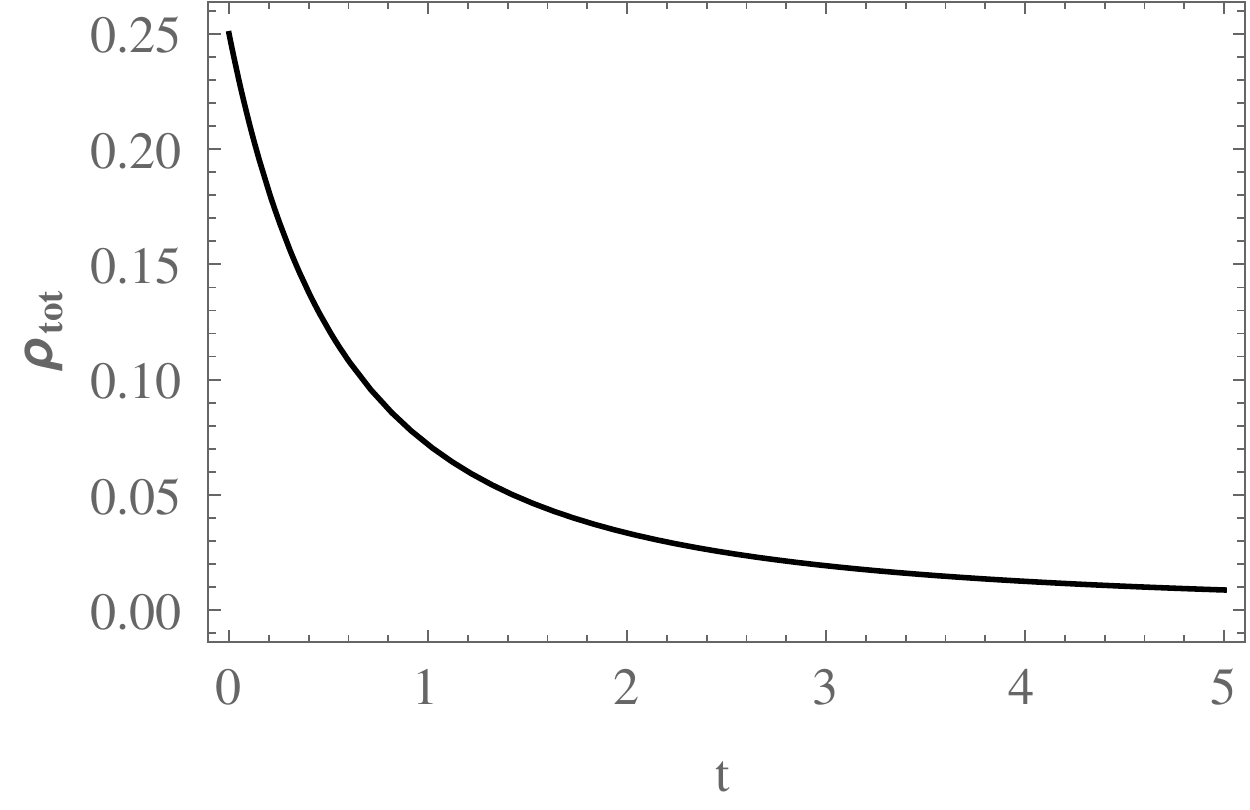}
\hspace{2mm}
\centering\includegraphics[width=3in]{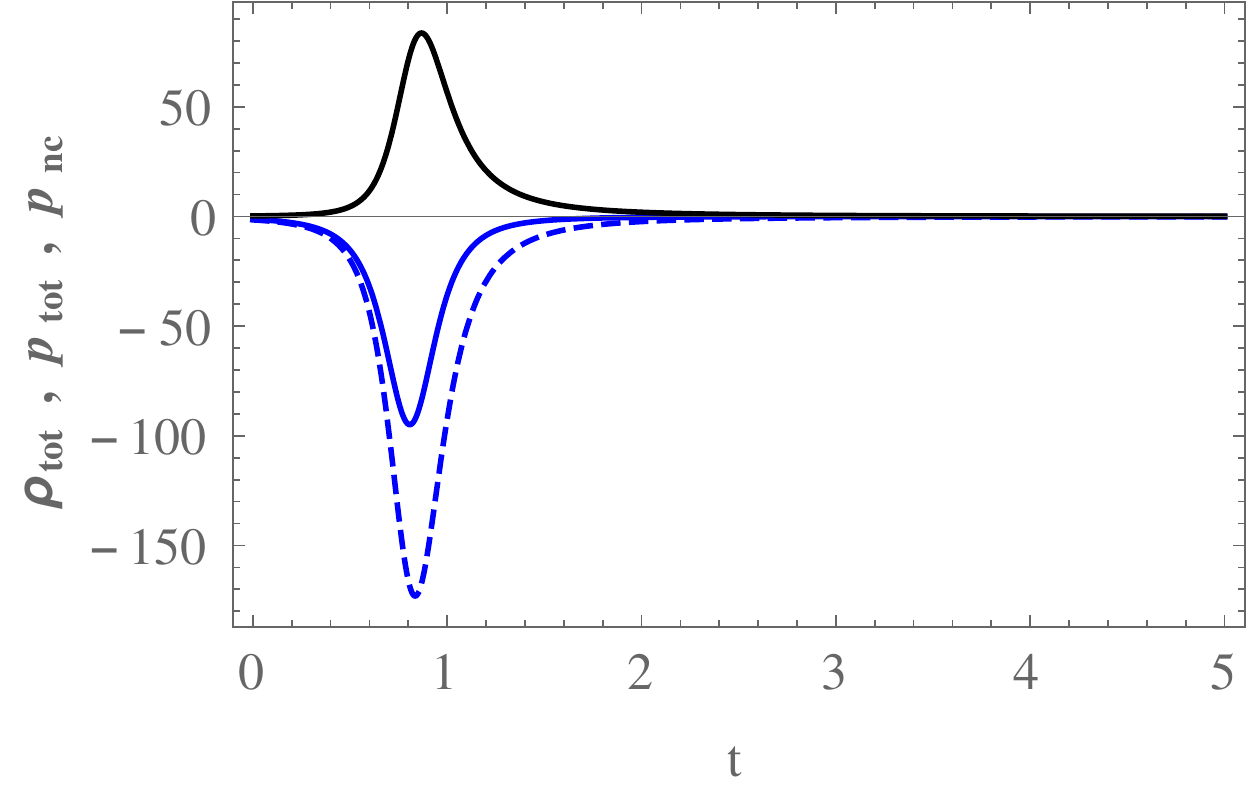}
\hspace{2mm}
\centering\includegraphics[width=3in]{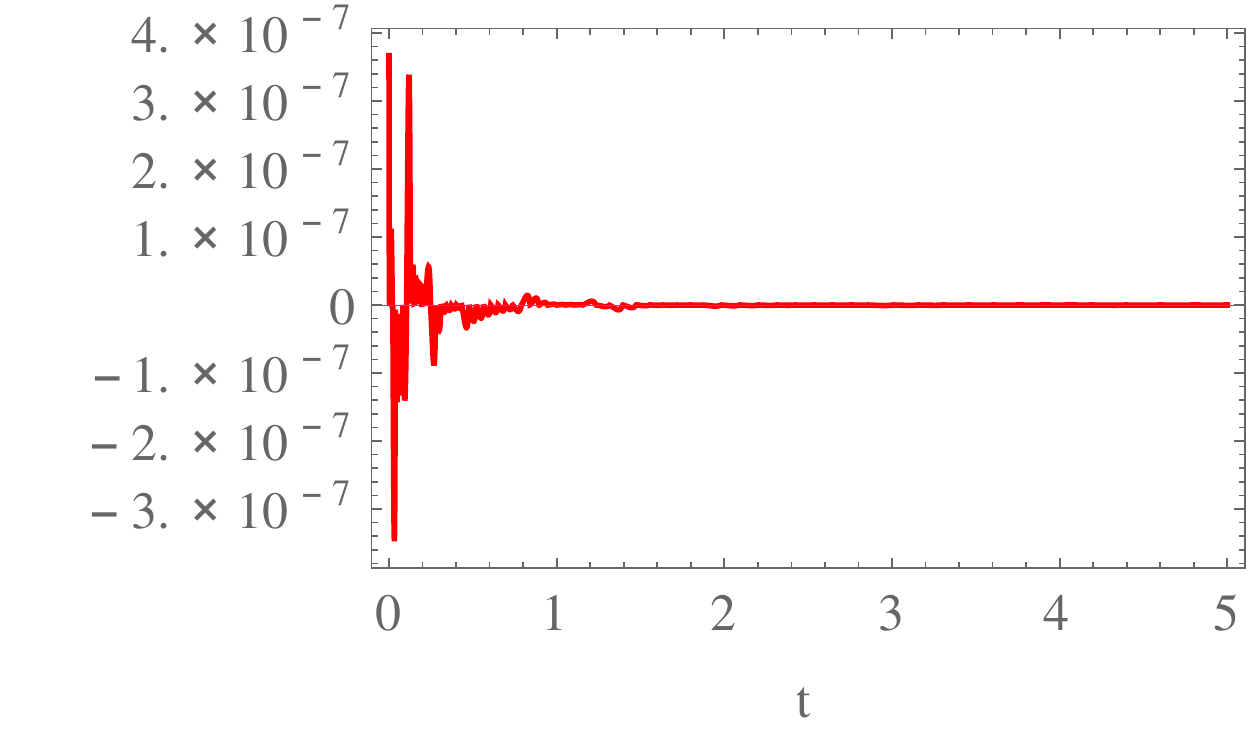}
\hspace{2mm}
\centering\includegraphics[width=3in]{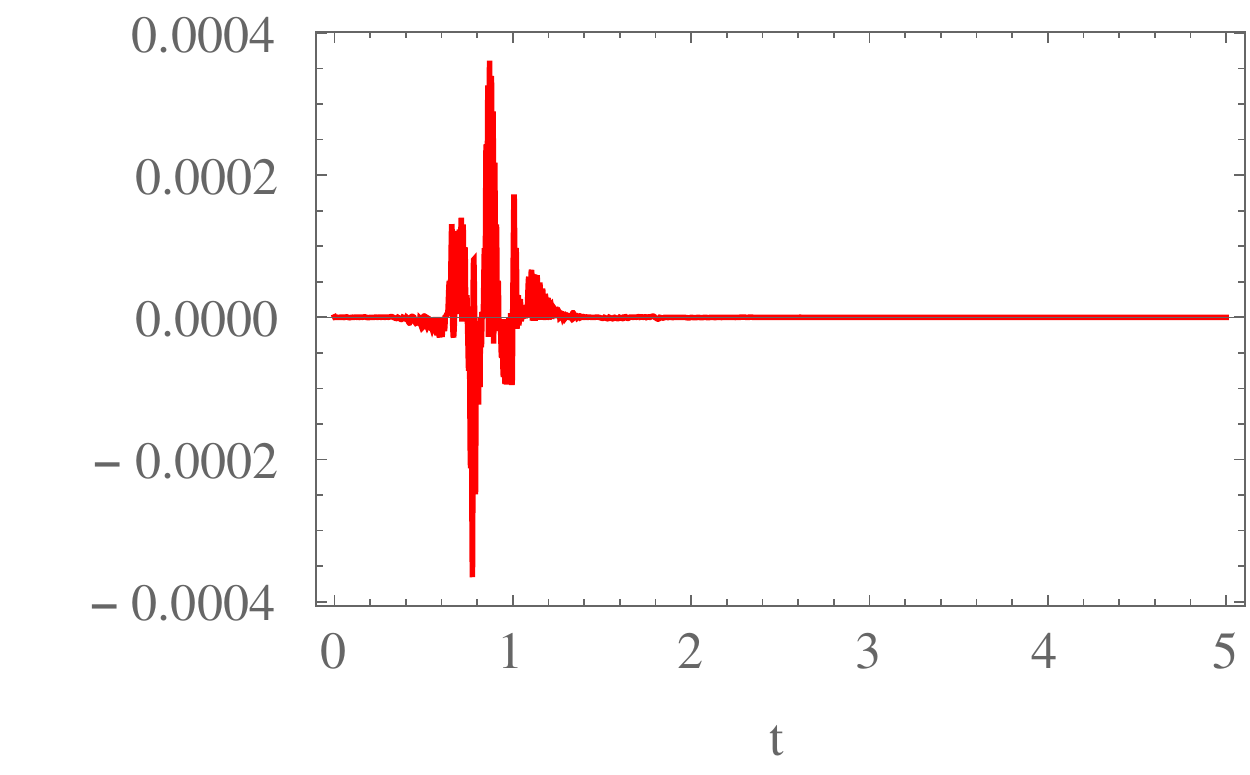}
\caption{The behavior of the total kinetic energy (black curves),
total pressure (solid blue curve)
and NC pressure (dashed blue curve) against cosmic time
for the commutative case (upper left panel) and
NC case (upper right panel).
Moreover, the lower left panel and
lower right panel are shown the numerical error concerning
satisfying the conservation equation \eqref{cons-eq} for
the commutative case and NC case, respectively.
The values for the initial conditions, parameters and units are
equal to the corresponding ones chosen for the figure \ref{phi-zero}.}
\label{rho-p}
\end{figure}

\begin{figure}
\includegraphics[width=3in]{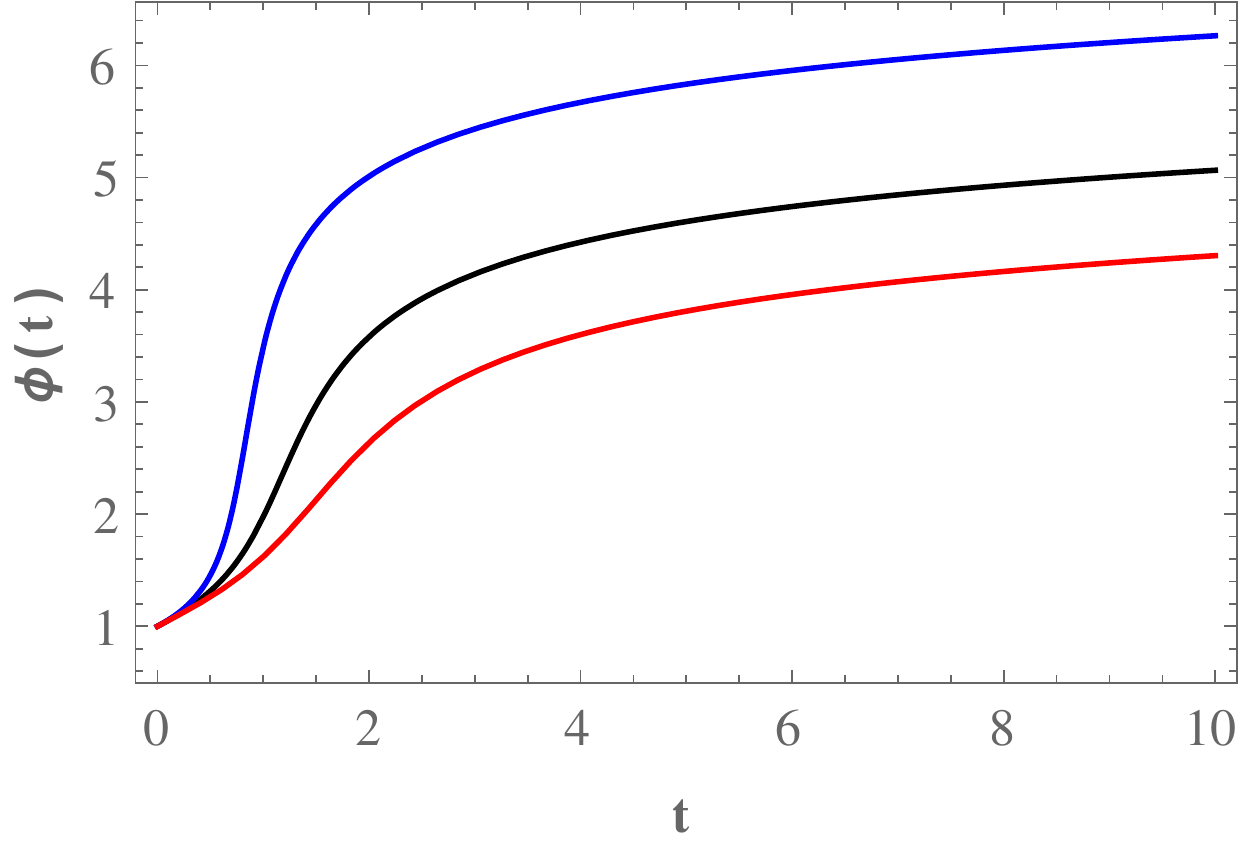}
\hspace{2mm}
\includegraphics[width=3in]{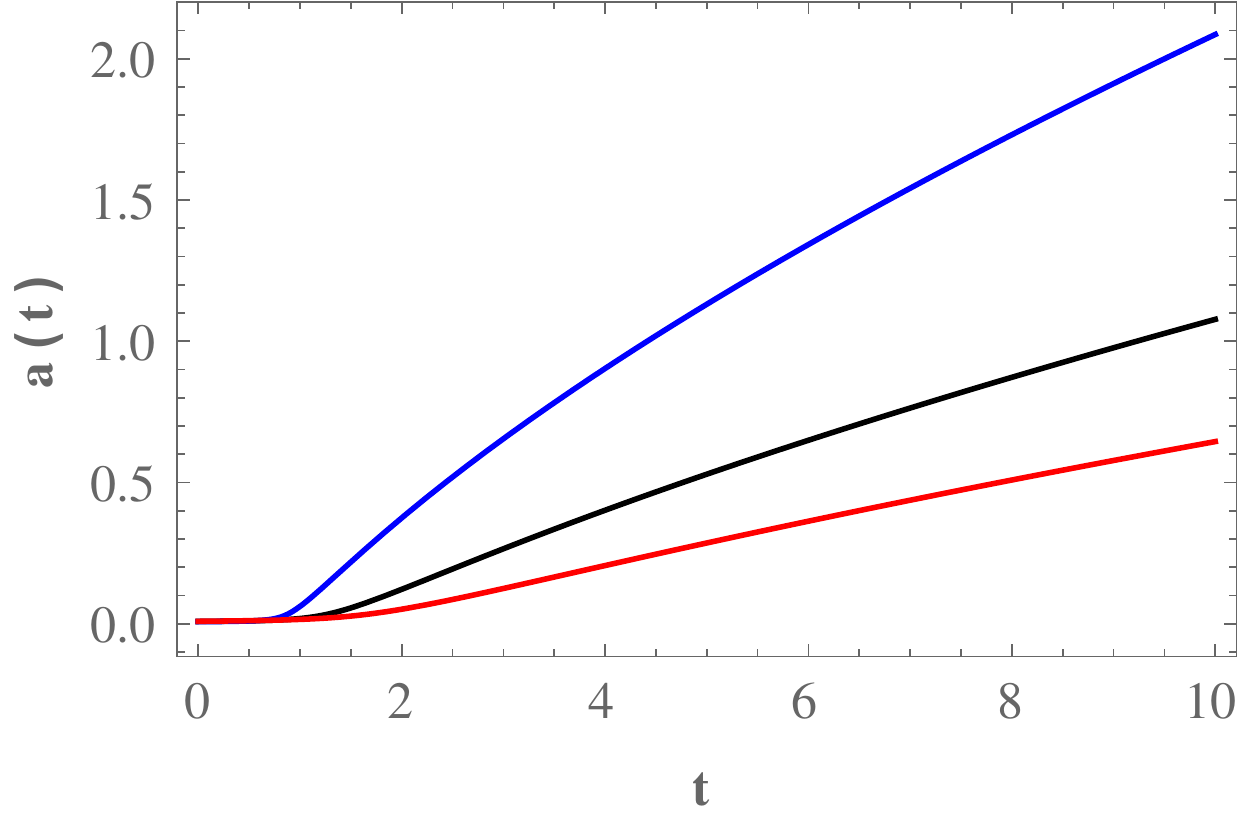}
\caption{The behavior of the scalar field $\phi(t)$ and the scale factor $a(t)$ against cosmic time
for our herein NC model with different values of ${\cal W}$:
The blue, black and red curves are associated with ${\cal W}=2$,
${\cal W}=3$ and ${\cal W}=4$, respectively.
Except ${\cal W}$ (which varies here), the values
for the initial conditions, other parameters and units are
equal to the corresponding ones chosen for the figure \ref{phi-zero}.}
\label{phi-diff-W}
\end{figure}

\begin{figure}
\includegraphics[width=3in]{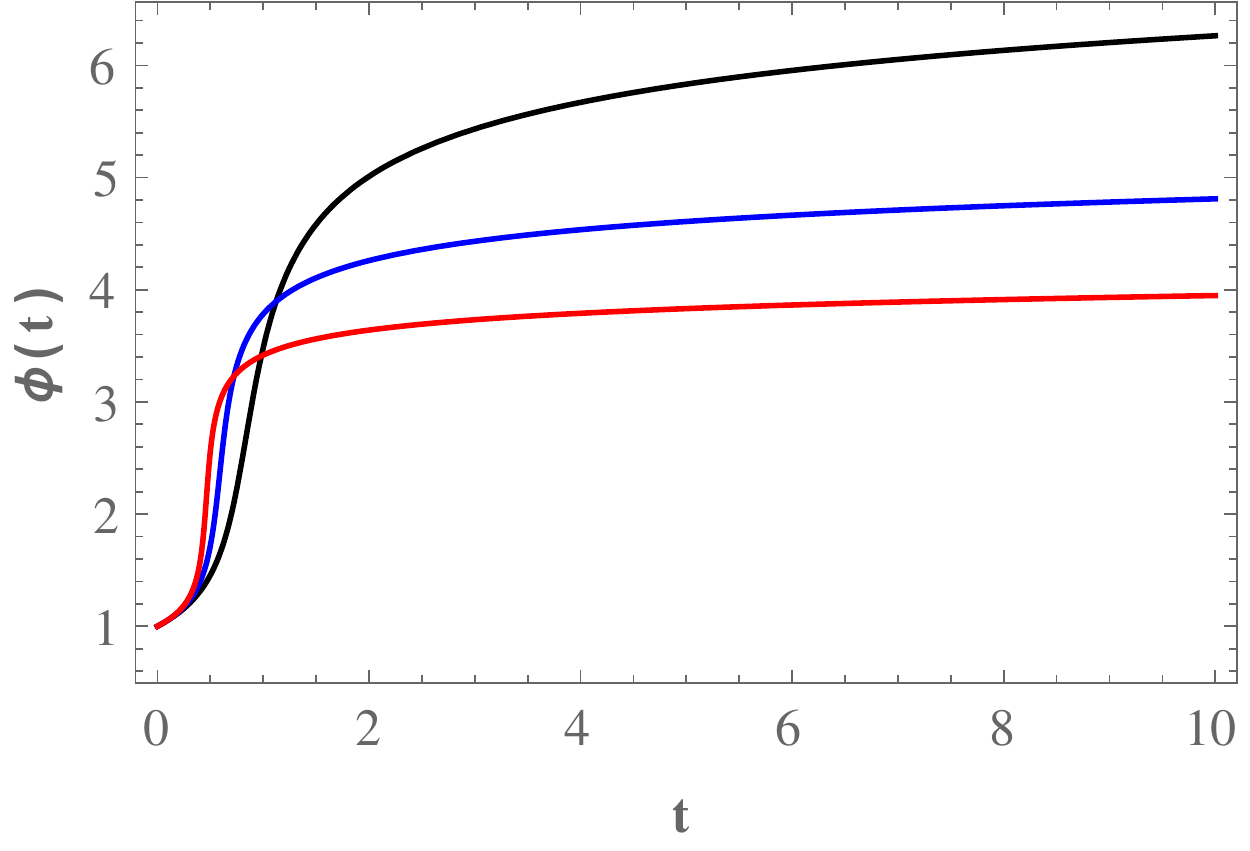}
\hspace{2mm}
\includegraphics[width=3in]{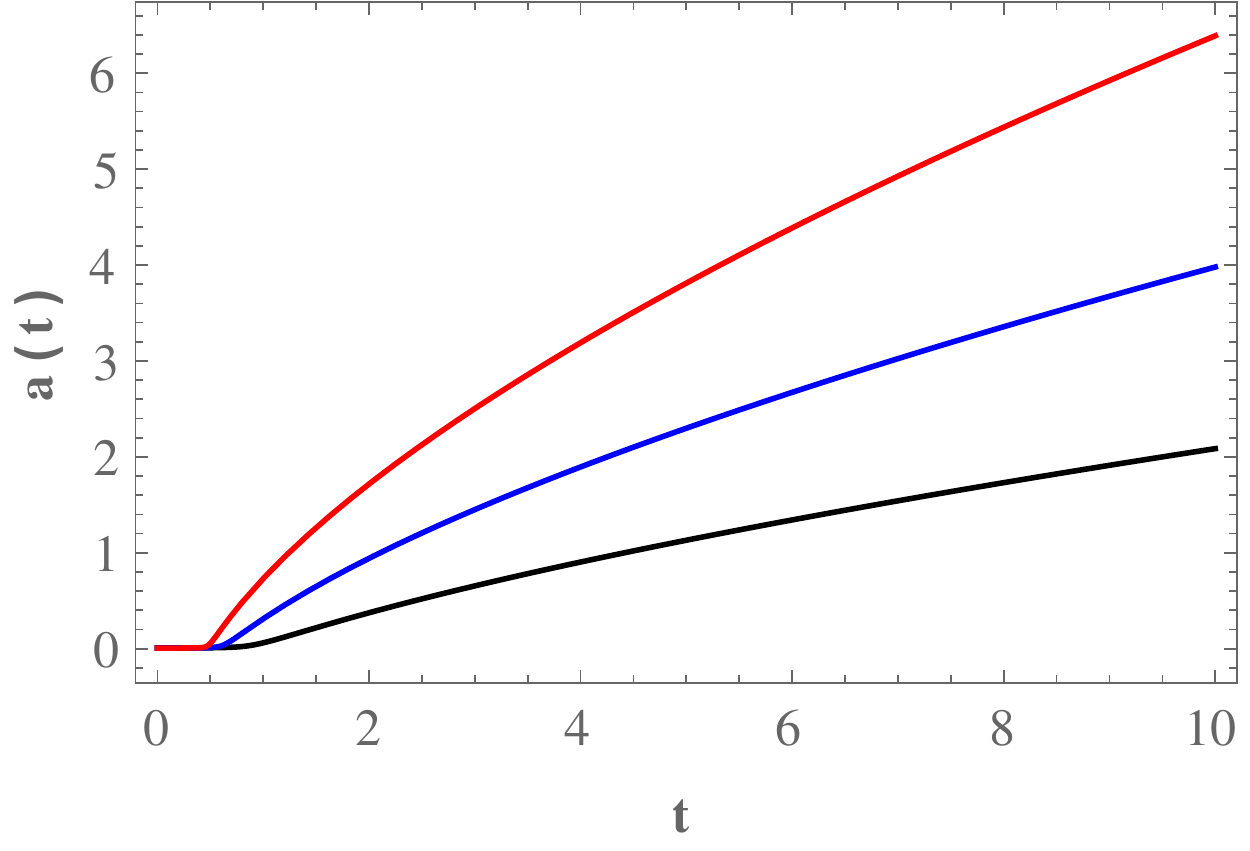}
\caption{The behavior of the scalar field $\phi(t)$ and the scale factor $a(t)$ against cosmic time
for our herein NC model with different values of $n$: The black, blue and red curves are associated with $n=1$,
$n=2$ and $n=3$, respectively.
Except $n$ (which varies here), the values
for the initial conditions, other parameters and units are
equal to the corresponding ones chosen for the figure \ref{phi-zero}.}
\label{phi-diff-n}
\end{figure}

\item
 Up to now, we have seen that our herein NC model can provide
an accelerating phase at early times, and soon after
 the scale factor can gracefully exit
from that accelerating phase and enter to a
decelerating phase, which could be assigned to the radiation-dominated era.
Therefore, it seems that our model, disregarding the 60
e-fold duration, can be considered as a proper inflationary scenario.
Notwithstanding, it has been believed that among the
problems associated with the standard cosmology, the horizon
problem is the most important one, which should be
resolved by a successful inflationary scenario.
In this respect, let us investigate only a nominal
condition as the key to resolving the horizon problem \cite{Lev95,Lev95-2}:
\begin{eqnarray}
D_{\gamma}\equiv d_{\gamma}-H^{-1}>0,\label{d-hor-1}
\end{eqnarray}
where $d_{\gamma}$ denotes the distance a photon has traveled
\begin{eqnarray}\label{d-hor-2}
d_{\gamma}(t)\equiv a(t)\int_{t_i}^{t}\frac{d\tilde{t}}{\tilde{a}(\tilde{t})}.
\end{eqnarray}
In order to check satisfaction of the nominal condition \eqref{d-hor-1}, we first should obtain $D_{\gamma}$.
 In this respect, for our herein NC, we substitute
the relations associated with the scale factor from
relations \eqref{Ex1-Vphi} into \eqref{d-hor-2}. Therefore, we obtain an integration over
$d\tilde{t}$ with an unknown integrand (as a function of the scalar field), which, in turn,
is obtained from \eqref{c-zero-V-1}.
Moreover, we should also substitute the Hubble parameter
(which can be also obtained from $\phi$)
from \eqref{exp-rel-2} into \eqref{d-hor-1}.
Consequently, investigating the nominal condition \eqref{d-hor-1} for
our herein NC model is not possible unless we
obtain $\phi(t)$ from solving \eqref{c-zero-V-1}. However, as mentioned, for
the NC case, we have to apply numerical analysis.
Our numerical endeavors have shown that condition \eqref{d-hor-1} is
satisfied for every set of values that yield the above-mentioned stages 1 to 6, see, for instance, figure \ref{Dh}.

\begin{figure}
\centering\includegraphics[width=3.2in]{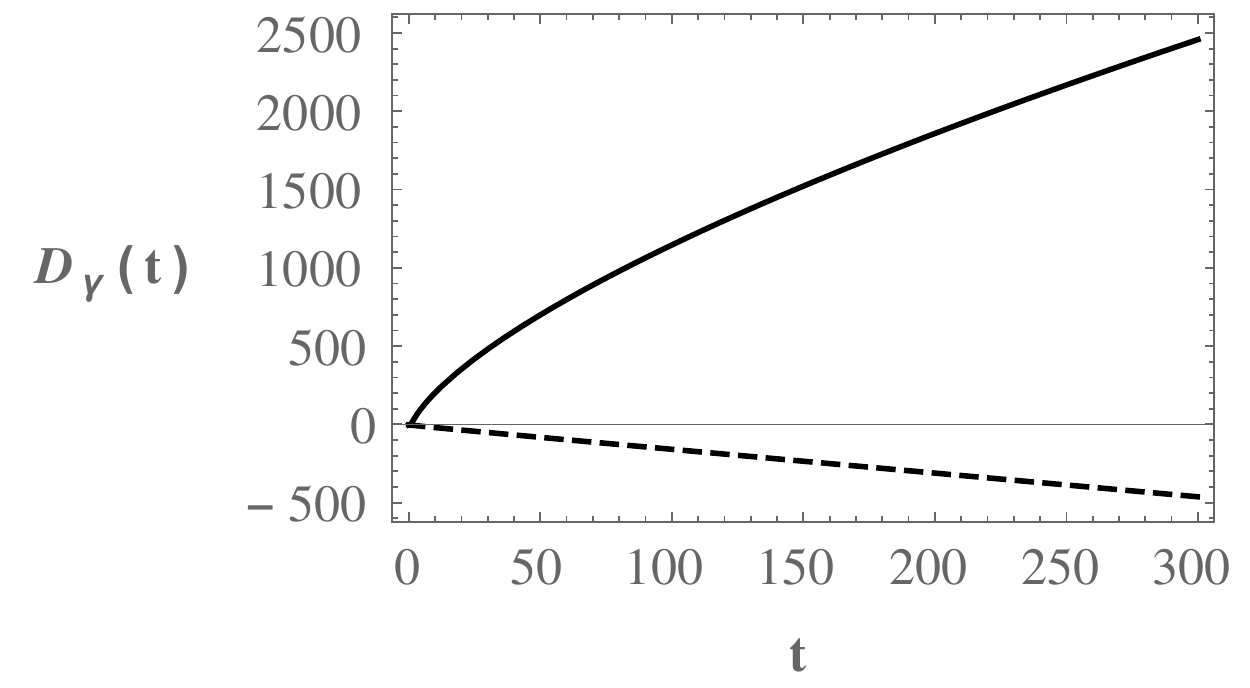}
\caption{The time behaviour of $D_{\gamma}$ associated with
 the commutative model (the dashed curve) and the NC model (the solid curve).
The values for the initial conditions, parameters of the model and units are
equal to the corresponding ones chosen for the figure \ref{phi-zero}. }
 \label{Dh}
\end{figure}

\end{enumerate}

\section{Cosmological dynamics in deformed phase scenario}
\label{dyn}
It seems that it is impossible to reconstruct
the Lagrangian of our NC model.
In this respect, let us focus on it at the level of the field equations.
Specifically, we can compare the evolution of the scale
factor for our herein NC cosmological setting, i.e.,
\begin{eqnarray}\nonumber
\frac{\dddot{a}}{a}&+&3\left(\frac{\dot{a}\ddot{a}}{a^2}\right)
-4\left(\frac{\dot{a}}{a}\right)^3+
\frac{\theta}{12\alpha a^2}\left[\phi_i^{\frac{n+2}{2}}+\frac{n+2}{2\alpha}\ln\left(\frac{a}{a_i}\right)\right]^3\\\nonumber\\
&\times&\Bigg\{\frac{\ddot{a}}{a}+3\left(\frac{\dot{a}}{a}\right)^2
\left(\frac{n+2}{2\alpha}\left[\phi_i^{\frac{n+2}{2}}+\frac{n+2}{2\alpha}\ln\left(\frac{a}{a_i}\right)\right]^{-1}-1\right)\Bigg\}=0,
\label{evol-a-2}
\end{eqnarray}
with that of the Starobinsky inflationary model \cite{S80}, see section \ref{concl}.
In order to obtain equation \eqref{evol-a-2}, we have used equations \eqref{asli2} and \eqref{Ex1-Vphi}.

Moreover, in order to confirm the results presented in
the previous section, let us provide an appropriate settings for the dynamical system.
In this regard, let us rewrite equation \eqref{evol-a-2} in more convenient form as
\begin{equation}
\dot{H}+3H^2+\frac{\theta H}{12\alpha a^2}\left[\phi_i^{\frac{n+2}{2}}+\frac{n+2}{2\alpha}\ln\left(\frac{a}{a_i}\right)\right]^3=0.
\label{H-tilde}
\end{equation}
Letting $y=\dot{a}$, then instead of equation (\ref{H-tilde}), we have
\begin{equation}
\dfrac{dy}{da}=-2\left(\frac{y}{a}\right)-\frac{\theta}{12\alpha a^2}
\left[\phi_i^{\frac{n+2}{2}}+\frac{n+2}{2\alpha}\ln\left(\frac{a}{a_i}\right)\right]^3\: ,
\label{y-dyn}
\end{equation}
which is very susceptible to the ICs, values of integration constants ($a_i$ and $\phi_i$)
and the parameters of the model, i.e., ${\cal W}$, $n$, and $\theta$.

\begin{figure}
\centering\includegraphics[width=3in]{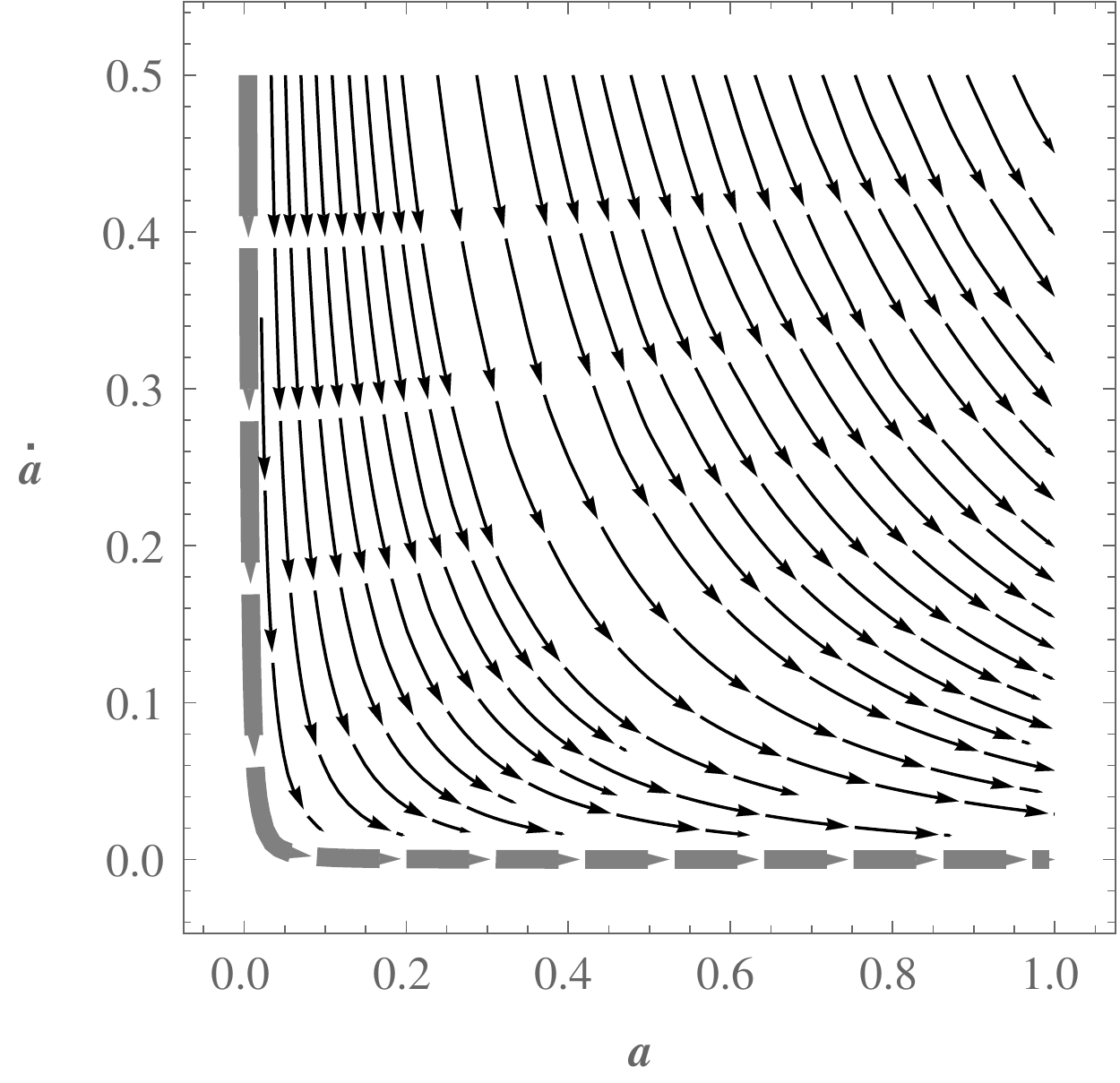}
\hspace{5mm}
\centering\includegraphics[width=3in]{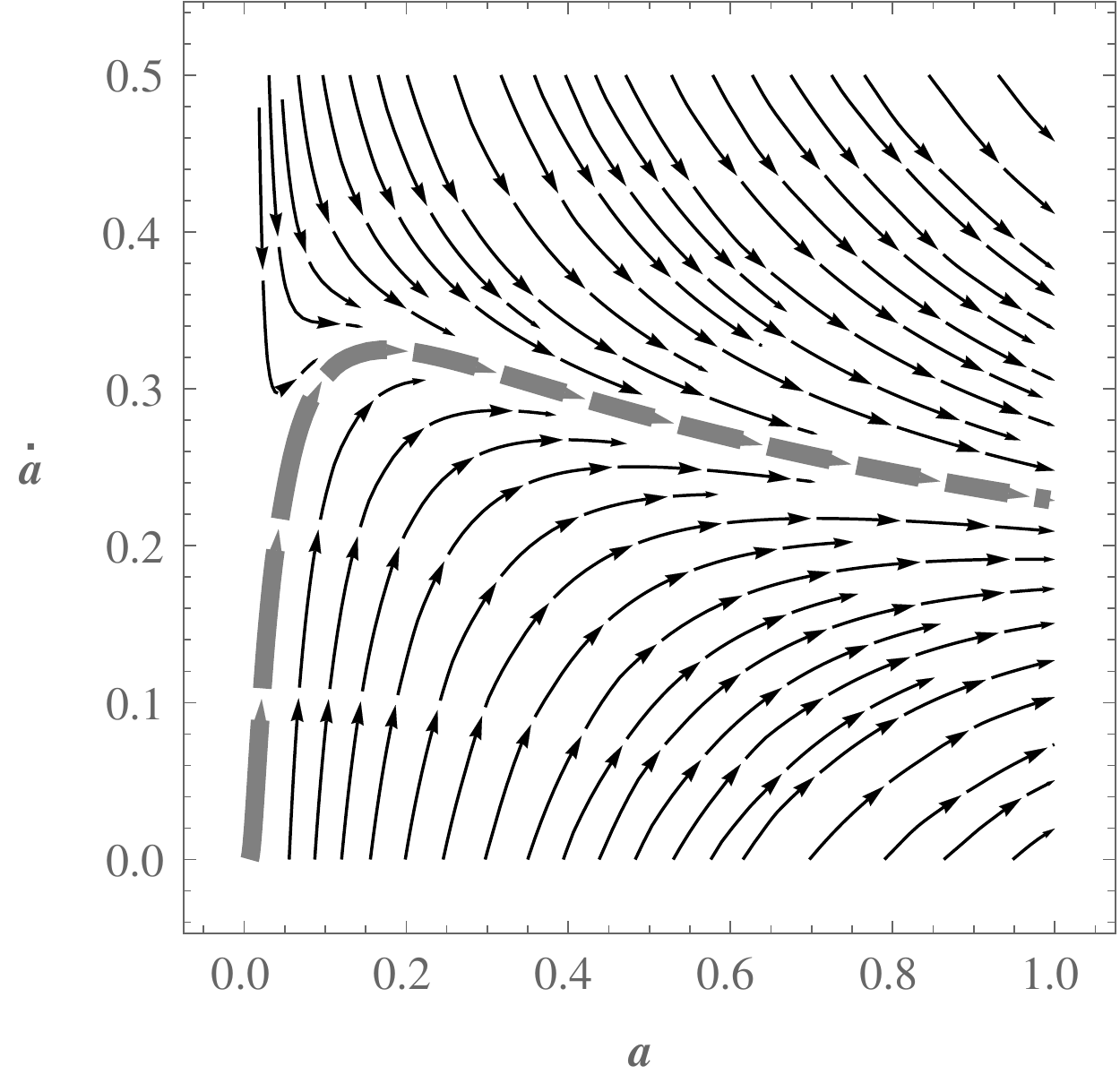}
\caption{Phase space portrait for equation (\ref{y-dyn}).
The left and right panels are associated with the commutative
case ($\theta=0$) and noncommutative case ($\theta=-0.001$), respectively.
We have also assumed $n=1$, ${\cal W}=2$, $8\pi G=1$, $a_i=0.005$ and $\phi_i=0.01$
 for both of the cases.}
\label{PhaseP}
\end{figure}

\begin{figure}
\centering\includegraphics[width=3in]{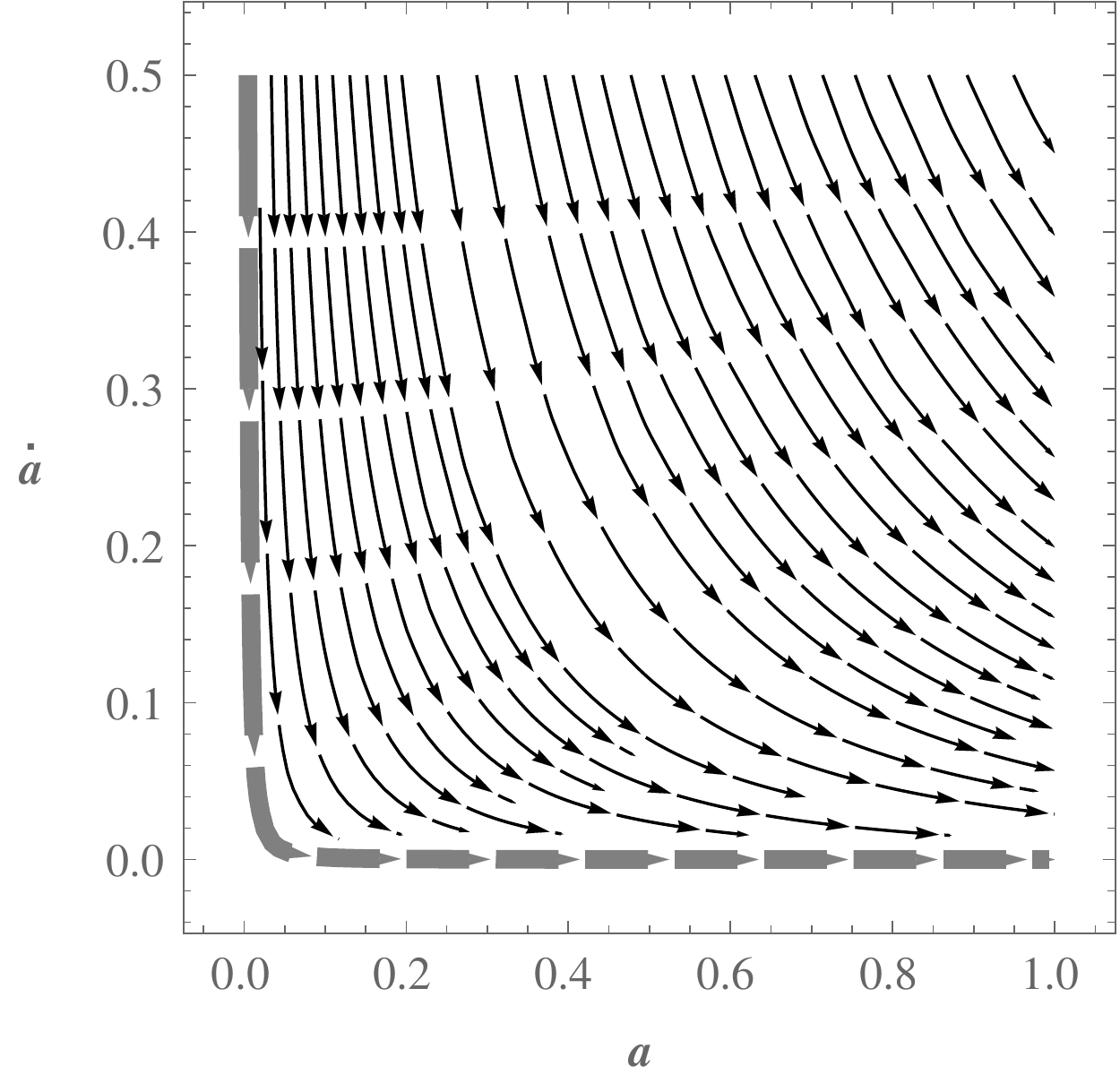}
\hspace{5mm}
\centering\includegraphics[width=3in]{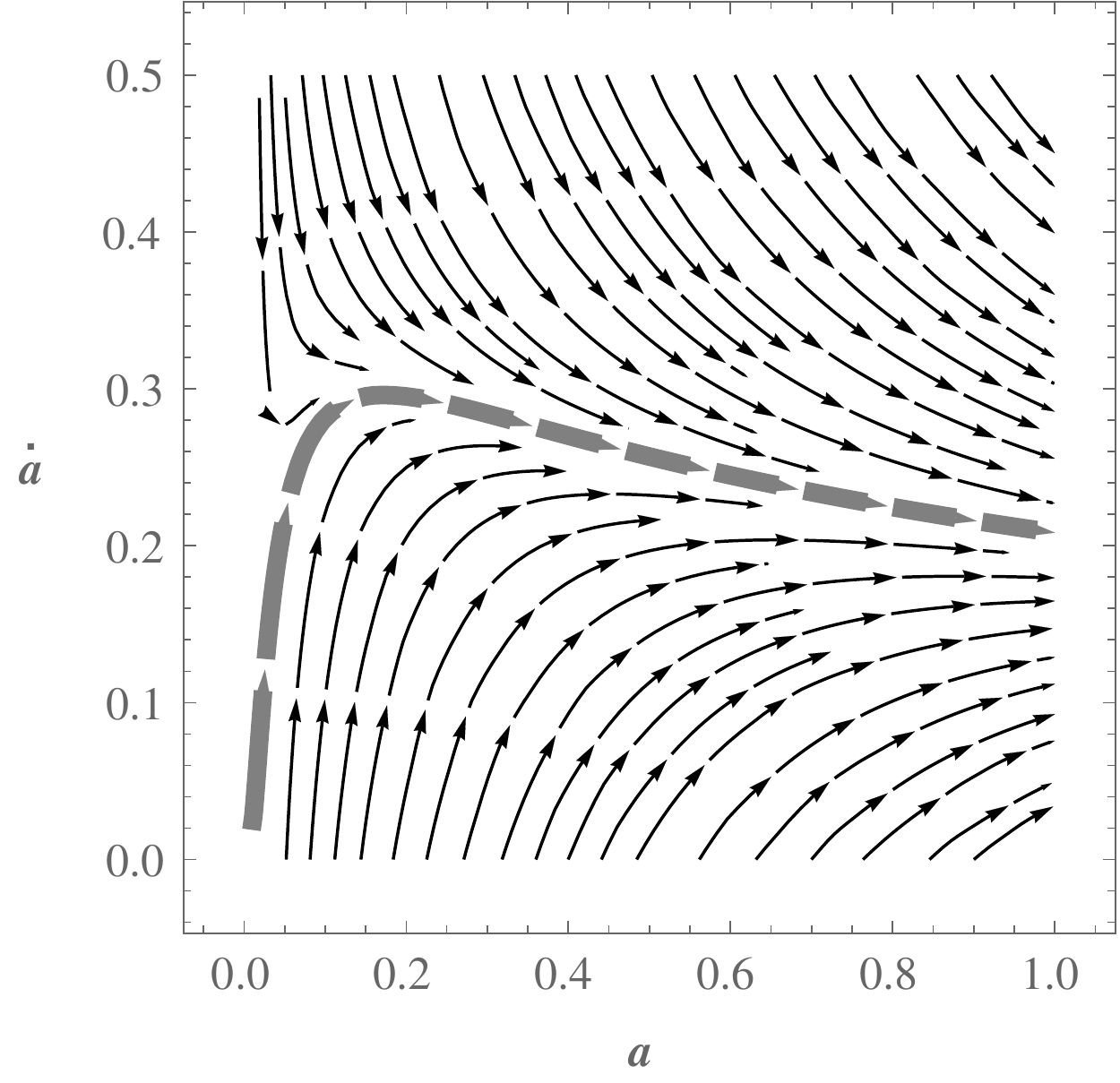}
\caption{Phase space portrait for equation (\ref{y-dyn}).
The left and right panels are associated with the commutative
case ($\theta=0$) and noncommutative case ($\theta=-0.0003$), respectively.
We have also assumed $n=6$, ${\cal W}=5$, $8\pi G=1$, $a_i=0.005$ and $\phi_i=0.01$
 for both of the cases.}
\label{PhaseP-new}
\end{figure}

Now, by plotting the phase portrait of equation (\ref{y-dyn}), the difference between the
commutative and noncommutative cases are clearly visible.
For the former (see the left panel of figure \ref{PhaseP}), for all the
solutions $(a, \dot{a})$, it is seen that $\dot{a}$ always decreases while $a$ increases.
Whereas, for the latter (see the right panel of the figure \ref{PhaseP}),
for very small values of the scale factor, we
observe an additional interesting behavior for all the
solutions $(a, \dot{a})$. More concretely, during a very short time, $\dot{a}$ increases until it
reaches its maximum value. Thereafter, it decreases
while the scale factor increases.
Finally, depending on the ICs and the values of the
parameters of the model, it gets constant values at late times.
It is worth noting that figure \ref{PhaseP} includes a vast
range of solutions such that the particular solution
shown in figure \ref{phi-zero} corresponds to one of
the trajectories plotted in figure \ref{PhaseP}.
This phase portrait, with more complete specifications, certifies
the inflationary phase (with a graceful exit) described in the previous section.
We should emphasize that our herein
NC model, for various sets of the parameters, can yield the
interesting results presented within the preceding section.
For instance, the figure \ref{PhaseP-new} shows the phase portrait of
equation (\ref{y-dyn}) with other values of $n$, ${\cal W}$ and $\theta$.
%In the particular case where $n=0$ and ${\cal W}=1$, from
%using \eqref{}, it is easy to show that equation \eqref{evol-a-2} reduced to the
%its counterpart obtained in \cite{RSFMM18}.

\section{Conclusions}
\indent \label{concl}

In this work, by considering the spatially
flat FLRW metric and applying the Hamiltonian formalism,
we first retrieved the equations of motion associated with a generalized SB theory.
Subsequently, by proposing a dynamical deformation
between the conjugate momenta (associated with the scale factor
and the SB scalar field), in which the deformation
parameter appears linearly, we have constructed a
noncommutative SB framework, such that in a particular case where $\theta=0$, the
commutative cosmological model is recovered (cf. Section \ref{Standard}).
%As, in our herein NC model, we focused on a special case where $n=0$,
%it is seen that letting ${\cal W}=0$ and assuming simple transformations, one easily retrieve
%the corresponding NC cosmological model constructed within
%the simplest minimally coupled canonical scalar field cosmology \cite{RSFMM18}.

In order to elucidate the kinetic acceleration
arisen from our herein NC model, we restricted our attention
to a simple case in which not only the
scalar potential but also the Lagrangian density associated with
the ordinary matter are absent. Then, we have shown that
it is straightforward to write the NC Klein-Gordon equation in terms
of the only SB scalar field and its time derivatives.
It is worthy to note that the NC parameter appears in that wave equation linearly, too.

Subsequently, to do the latter comparison, we first obtained an exact
cosmological solution for the standard case.
Whilst, concerning the NC model, we found that it is not
feasible to obtain an exact analytic solution for the complicated nonlinear
differential equation (i.e., the NC Klein-Gordon equation). Therefore, we
 resorted to applying numerical methods to analyze the time behavior of the physical quantities.

In contrast to the corresponding standard SB model, our numerical endeavors have
indicated that our herein simple NC model
(in which the NC parameter appears linearly not
only in the proposed deformed Poisson bracket but also in the
field equations) can provide
fascinating aspects. Let us be more precise.
We have shown that our NC model
yields a kinetic acceleration at early times.
Thereafter, the scale factor can exit gracefully from that phase
and enter a decelerating one.
The latter can be assigned to the radiation-dominated phase.
Therefore, we have interpreted this outcome, which is attained due to the presence of the NC effects,
as an inflationary phase for the universe at early times
and shown that its corresponding expected nominal condition is satisfied.
Moreover, at the late times, we also observed the effects of the noncommutativity:
we showed that the scale factor increases with constant
speed (presence of a zero acceleration epoch), which
can be assigned to a coarse-grained explanation.

We have also depicted the time behavior of the NC
energy density, NC pressure and then compared them
 with the corresponding counterparts associated with
the non-deformed model. We have shown that the time
behavior of the quantities depends not only on the NC parameter but also on the
values taken by the parameters ${\cal W}$ and $n$, as expected.

Finally, we constructed an appropriate dynamical setting, by
which we easily illustrated the effect of the noncommutativity.
More concretely, using the same set of ICs and the
parameters of the model applied for plotting figure \ref{phi-zero}, we have depicted the
phase portrait of equation (\ref{y-dyn}). It is seen that the sole trajectory
of that plot corresponds to those of the figure
\ref{phi-zero} (for either commutative model or NC model),
which confirms all the results of subsection \ref{NC case}.

Before closing this section, it is worthy to mention a few comments regarding the strengths and
shortcomings of our herein NC model:
\begin{itemize}
\item
In this work, we have investigated
the effects of the noncommutativity for a particular case.
More concretely, we have restricted our attention to a
special case where (i) the ordinary matter and the scalar potential are absent; (ii) a particular
dynamical deformation between only the conjugate momenta was proposed.
Obviously, by removing either one or more of the above
restrictions, one can construct more extended models, which may yield more interesting results.
For instance, generalizing this work to a NC model
including a non-vanishing scalar potential, but still admitting
the other constraints, we can establish
NC counterparts for the deformed versions of Luccin-Mataresse model \cite{LM85} and
Barrow-Burd-Lancaster-Madsen model \cite{BBLM86, GC07}, see also \cite{GC07}.
The generalized version of the former and of the latter
 has been established in the non-deformed phase space
 in the context of SB theory \cite{RSM22}.
 Such extended frameworks have been investigated and will be presented within our forthcoming works.

\item
In comparison with the NC model presented in \cite{RSFMM18}, we observe that in
our herein NC model, there are two extra free
parameters, i.e., ${\cal W}$ and $n$, by which one can not only provide
different behaviors for the physical quantities
but also it may assist to retrieve appropriate values
for the e-fold number (which is also one of the
essential features of a expected inflationary epoch)
to be in agreement with the observational data.
In a particular case where ${\cal W}=1$ and $n=0$, we recover
the corresponding model investigated in \cite{RSFMM18}.
Moreover, in another particular case
where $\varrho(a)=0$ and $n=-2$, using transformation
\begin{equation}
\phi=\varphi_i\exp\left(\frac{1}{\sqrt{{\cal W}}}\frac{\varphi}{\varphi_i}\right),
\label{phi-transf}
\end{equation}
where $\varphi_i={\rm constant}$ and $\varphi$ carry the same dimension of $\phi$, the
Lagrangian \eqref{eq1} transforms to the corresponding standard
minimally coupled scalar field.
Therefore, in the commutative case, we recover the model studied in \cite{RSFMM18}.
However, it is important to note that, under the
transformation \eqref{phi-transf}, the deformation \eqref{deformed} is not equivalent to that chosen in \cite{RSFMM18}.
Concretely, the NC case corresponding to $n=-2$ will be different from that investigated in \cite{RSFMM18}.
This particular kinetic model was contributed to other extended NC models mentioned in the preceding comment.
\item
We should emphasize that it is almost impossible to
retrieve the Lagrangian associated with our NC
model, and therefore, it is a complicated
procedure to investigate the quantum features
of the model by means of the perturbation analysis.
In this respect, at the level of the field equations, we
have obtained a proper NC differential equation associated with the evolution of the scale factor.
By means of such a procedure as well as by establishing the
corresponding dynamical setting, one may probe
the possible relations between the parameters appeared in our model (i.e.,  NC
parameter, SB coupling parameter, $n$ and the
integration constants)
and the quantum corrections observed in the Starobinsky
inflationary model to find a feasible correspondence between these scenarios.
\end{itemize}

\section*{Acknowledgments}
I would thank the anonymous reviewers for critical reading of this
manuscript and their valuable comments.
I acknowledge the FCT grants UID-B-MAT/00212/2020 and UID-P-MAT/00212/2020 at CMA-UBI plus
the COST Action CA18108 (Quantum gravity phenomenology in the multi-me\-ssen\-ger approach).
%I do appreciate the
%organizers of the conference ALTECOSMOFUN'21, where a very particular case of this manuscript was presented.
%%%%%%%%%%%%%%%%%%%%%%%%%
%\end{paracol}

%\end{adjustwidth}
\end{document}